# Atomic diffusion-induced polarization and superconductivity in topological insulator-based heterostructures

Xian-Kui Wei[1]*†, Abdur Rehman Jalil[2], Philipp Rüßmann[3,4]*, Yoichi Ando[5], Detlev Grützmacher[2], Stefan Blügel[4], Joachim Mayer[1,6]

[1]Ernst Ruska-Centre for Microscopy and Spectroscopy with Electrons, Forschungszentrum Jülich GmbH, 52425 Jülich, Germany

[2]Peter Grünberg Institute and JARA-FIT, Forschungszentrum Jülich GmbH, 52425 Jülich, Germany.

[3]Institute for Theoretical Physics and Astrophysics, University of Würzburg, 97074 Würzburg, Germany.

[4]Peter Grünberg Institute and Institute for Advanced Simulation, Forschungszentrum Jülich GmbH and JARA, 52425 Jülich, Germany.

[5]Physics Institute II, University of Cologne, Zülpicher Str. 77, 50937 Köln, Germany

[6]Central Facility for Electron Microscopy, RWTH Aachen University, Ahornstraße 55, 52074 Aachen, Germany.

*Corresponding author Emails: xkwei@xmu.edu.cn; p.ruessmann@fz-juelich.de

†Present address: College of Chemistry and Chemical Engineering, Xiamen University, 361005, Xiamen, China

## ABSTRACT

The proximity effect at a highly transparent interface of an *s*-wave superconductor (S) and a topological insulator (TI) provides a promising platform to create Majorana zero modes in artificially designed heterostructures. However, structural and chemical issues pertinent to such interfaces are poorly explored so far. Here, we report the discovery of Pd diffusion induced polarization at interfaces between superconductive $Pd_{1+x}(Bi_{0.4}Te_{0.6})_2$ (xPBT, $0 \leq x \leq 1$) and Pd-intercalated $Bi_2Te_3$ by using atomic-resolution scanning transmission electron microscopy. Our quantitative image analysis reveals that nanoscale lattice strain and QL polarity synergistically suppress and promote the Pd diffusion at the normal and parallel interfaces, formed between Te-Pd-Bi triple layers (TLs) and Te-Bi-Te-Bi-Te quintuple layers (QLs), respectively. Further, our first-principles calculations unveil that the superconductivity of xPBT phase and topological nature of Pd-intercalated $Bi_2Te_3$ phase are robust against the broken inversion symmetry. These findings point out the necessity of considering coexistence of electric polarization with superconductivity and topology in such S-TI systems.

Keywords: topological insulator, superconductivity, polarization, atomic diffusion and intercalation, scanning transmission electron microscopy

**INTRODUCTION**

Majorana zero modes (MZMs) are one of the most exciting research topics in condensed matter systems owing to their potential applications in quantum computation.[1-3] In essence, MZMs obey non-Abelian braiding statistics.[4] Under unitary gate operation, the nonlocal encoding of the quasiparticle state makes the computation immune to certain type of error caused by local perturbation, thus leading to fault-tolerant computation.[5,6] Relying on non-trivial topological states of matter, the MZMs are predicted to emerge either in spinless *p*-wave topological superconductors (Ss) with one or two dimension,[7] or via proximity-induced superconductivity at interfaces of *s*-wave Ss with topological insulators (TIs).[8] Meanwhile, possible signatures of MZMs have also been unveiled in a number of systems like semiconductors,[9] quantum anomalous Hall insulators[10] and magnetic atomic chains[11] such as InSb-NbTiN,[12] Fe-Pb,[11] $FeTe_{0.55}Se_{0.45}$ superconductors,[13] EuS-Au[14] and LiFeAs.[15]

Given that topological Ss are scarce, implementing the interface-based proximity effect becomes a natural choice to pursue the MZMs and to construct heterostructure-based devices. Among various TIs such as HgTe, BiSb and $PbBi_2Te_4$,[16-18] 2D van der Waals (vdW) layered $(Bi_{1-x}Sb_x)_2Te_3$ has been widely investigated for its tunable topological surface state by chemical doping[19-21] and control of growth condition.[22,23] In spite of the ease in fabricating S-TI nanostructures, e.g., using stencil-lithography-based molecular-beam epitaxy,[24,25] such hybrid devices usually suffer issues about chemical diffusion, electronic structure change and interfacial dipole layers.[26,27] Therefore, clarifying the elemental diffusion mechanism and the fundamental physical properties at the interface becomes the urgent task towards creating stable MZMs in such hybrid S-TI devices.

Lately, $PdTe_2$-based Ss receive considerable attention owing to their intriguing band structure and transport property. Studies report that pure $PdTe_2$ is a Dirac semimetal with a superconducting $T_C$ around 1.64 K.[28,29] By increasing the concentration of Pd, the $T_C$ of $Pd_{1+x}Te_2$ (x ≥ 0) can be increased to 4.5 K in metallic PdTe.[30,31] Although a potential phase boundary is expected in the structure-composition phase diagram,[31] a continuous $PdTe_2$-to-PdTe solid solution via gradual addition of Pd at the vdW gaps seems to refutes the existence of the boundary (Fig. 1A). Intriguingly, when metallic Pd is deposited on a TI like $Bi_2Te_3$ (Fig. 1B), a $PdTe_2$-like superconducting phase ($T_C \approx 0.6$ K) spontaneously forms at the interface through diffusing Pd into the TI.[32] The newly formed fresh S/TI interface offers an alternative approach to create MZMs via the proximity effect.[33,34] Other than this, it has been claimed that Pd diffusion into $Bi_2Te_3$ can also lead to formation of a superconductive phase,[35,36] which indicates controversy about the origin of superconductivity.

In this work, in order to unravel the diffusion-based fundamental physics and device application, the diffusion pathway of Pd into the $Bi_2Te_3$ films grown on Si (111) substrates is investigated by using atomic-resolution scanning transmission electron microscopy (STEM) Unexpectedly, associated with observation of chemical intercalation, the Pd diffusion induced polarization is observed in the intermediate $Pd_{1+x}(Bi_{0.4}Te_{0.6})_2$ (xPBT, $0 \leq x \leq 1$) phase and at the xPBT/$Bi_2Te_3$ interfaces, i.e., at the normal vs parallel contact interfaces. Apart from disclosing the Pd diffusion pathway, our first-principles calculations reveal robustness of superconductivity for the xPBT and topology for the Pd-intercalated $Bi_2Te_3$ against the broken inversion symmetry. These findings highlight the necessity of exploring polarization-superconductivity-topology coupling in such S-TI systems.

**RESULTS AND DISCUSSION**

In our experiments, molecular beam epitaxy is used to grow the $Bi_2Te_3$ films (~18 nm) and a Pd layer (~6 nm) is deposited on top of $Bi_2Te_3$ to construct the S-TI heterostructures. Distinct from the Nb capped case,[24] our high-angle annular-dark field (HAADF) STEM imaging reveals that the capped Pd undergoes a spontaneous diffusion into the $Bi_2Te_3$, leading to formation of an intermediate $Pd_{1+x}(Bi_{0.4}Te_{0.6})_2$ (xPBT) phase between the Pd surface layer and the $Bi_2Te_3$ film (Fig. 1C). Although the Pd penetration depth varies according to the synthetic conditions, e.g., the substrate temperature during metal deposition,[33] thickness of the xPBT phase is observed to vary in the range of 6.8 to 8.6 nm in this specific case. As for roughness of the xPBT/$Bi_2Te_3$ interface, our energy dispersive X-ray spectroscopy (EDS) data reveals that this is attributed to quintuple layer (QL) terraces resulting from varied Pd diffusion depth into the TI (Fig. 1D).

The medium-resolution HAADF image shows that the xPBT phase is characteristic of a mixture of the $PdTe_2$- and PdTe-like triple layers (TLs), which consists of parallel and zigzag-type TLs due to inhomogeneous distribution of Pd atoms at the vdW gaps. In structure, the $PdTe_2$- and PdTe-like phases differ mainly in null and full occupancy of intercalated Pd atoms within the vdW gaps[29] (Fig. 1A). This is substantiated by reproduction of the PdTe phase, previously determined to have *P6$_3$/mmc* space group,[30] via adding one Pd atom at (0.0, 0.0, 0.5) site in the PdTe$_2$ (space group *P-3m*) framework (Supplementary Fig. S1). This also supports absence of the PdTe$_2$-PdTe phase boundary in the Pd$_{1+x}$Te$_2$ (0 ≤ x ≤ 1) phase diagram (Fig. 1A). One should note that the structure feature of the xPBT phase differs drastically from that of superconductive PdBiTe,[37] which has an intrinsic polar space group P2$_1$3.

Further, we observe two kinds of interfaces near the QL terraces, a normal interface and a parallel one, which are defined in terms of crystal plane orientation (yellow and green dashed lines) in the TLs and QLs (Fig. 1E). In combination with geometric phase analysis (GPA),[38] we find that the *a*-axis difference between the xPBT and Bi$_2$Te$_3$ phases gives rise to an array of interfacial mismatch dislocations (average spacing ~7 nm), as manifested by in-plane $\varepsilon_{xx}$ strain map (Fig. 1F). This is similar to mismatch dislocations observed at heterointerfaces of three-dimensional oxides.[39] Specifically, an out-of-plane $\varepsilon_{zz}$ strain map differentiates the two contact interfaces. At the normal interface, the *c*-axis expansion is highly condensed at the first two TLs (width ~ 1.7 nm) with $\varepsilon_{zz-max} \approx 0.193$ (Fig. 1G,H). While at the parallel interface, the lattice expansion extends to about four TLs (width ~ 3.5 nm) with $\varepsilon_{zz-max} \approx 0.066$. By amplifying the $\varepsilon_{zz}$ map, we see further details about *z*-direction lattice mismatch (Fig. 1I-K). Irrelevant to the interfacial contact manner, the local lattices (radius ~ 0.1 nm) on the left and right side of the dislocation cores are expanded and compressed, respectively. Our image analysis reveals that this arises from asymmetric agglomeration of Pd atoms near the dislocation cores (see Supplementary Fig. S2).

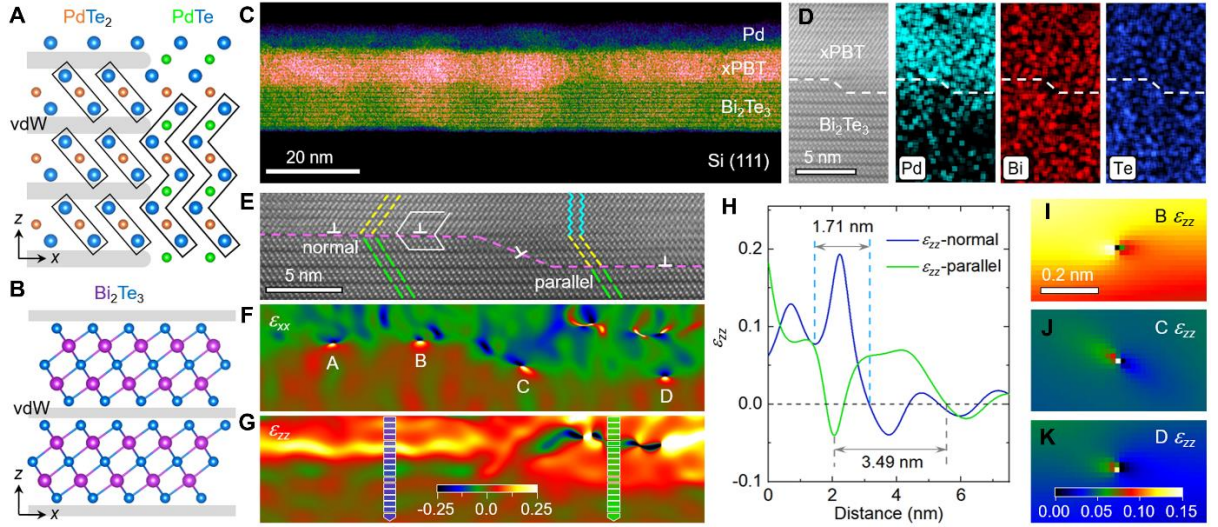

**Figure 1. Intermediate xPBT phase and interfacial mismatch dislocations.** (**A,B**) Crystal structures of PdTe$_2$, PdTe and Bi$_2$Te$_3$ viewed along [100] direction, respectively. The vdW gaps are denoted by grey stripes. (**C**) Low-magnification HAADF STEM image of a Pd/xPBT/Bi$_2$Te$_3$ heterostructure grown on Si (111) substrate. (**D**) HAADF image of a curved interface region and corresponding elemental EDX maps of Pd, Bi and Te. (**E-G**) Medium-magnification HAADF image of the interface and GPA strain maps of $\varepsilon_{xx}$ and $\varepsilon_{zz}$, respectively. The burgers vector of the mismatch dislocations is $\vec{b}$ = **a** [100] on the flat (001) planes. The yellow, cyan, and green line segments denote the TLs and QLs, respectively. (**H**) Local strain line profile of $\varepsilon_{zz}$ extracted from the normal and parallel interfaces illustrated in (**G**). (**I-K**) Magnified $\varepsilon_{zz}$ strain maps near dislocations labeled by B, C and D in (**F**), respectively.

As for the Pd self-diffusion induced xPBT phase, the HAADF image contrast, proportional to $Z^{1.7}$ ($Z$, atomic number),[40] indicates that the intercalated Pd atoms at the vdW gaps exhibit an irregular occupancy between the TLs (Fig. 2A,C and Supplementary Fig. S3). This leads to nanoscale bending of the TLs (white dashed lines) and possible presence of flexoelectricity, i.e., coupling of strain gradient with polarization or vice versa. By measuring positions of atomic columns via 2D Gaussian functional fitting,[41] our mapping reveals a short-range Pd displacement order relative to centers of nearest-neighboring Bi/Te columns ($\delta_{Pd-Bi/Te}$) in the TLs (yellow arrows). On the other hand, the Pd atoms at vdW gaps between the TLs tend to exhibit an opposite displacement order (blue arrows). This gives rise to an oscillating polar feature as manifested by line profiles of $\delta z_{Pd-Bi/Te}$ and $\delta x_{Pd-Bi/Te}$, which are averaged along x direction (Fig. 2B). Specifically, the nanoscale ordering of Bi and Te atoms along z direction breaks the structural inversion symmetry and thus leads to emergence of an intrinsic polar order. This is supported by non-negligible charge transfer from Bi/Te to Pd atoms owing to their difference in electronegativity, Bi ($\chi$ = 2.02), Te (2.10) and Pd (2.20), as verified in nickel phosphides [42].

Near the xPBT/Bi$_2$Te$_3$ interface, the HAADF images show that a mixture of parallel and zigzag TLs dominate on the xPBT side, which is attributed to fractional occupancy of Pd at the vdW gaps (Fig. 2D,F). Nevertheless, more Pd atoms are observed to diffuse into the Bi$_2$Te$_3$ across the parallel interface, which is evidenced by more Pd atoms locating at the interstitial sites within the QLs (green circles). This reveals that the intermediate xPBT phase originates from dismembering the Bi-Te bonds of the QLs through the interstitial Pd atoms. On this basis, different dismemberment processes can be proposed to understand the Pd diffusion pathways at the two interfaces (see Supplementary Fig. S4). By measuring the interplanar spacing ($d_{\text{IPS}}$-c), we find that the $a$ axis undergoes a sharp transition near the normal interface, while a gradual evolution is observed at the parallel interface (Fig. 2E,G). With respect to the average vdW spacing, $d_0 = 0.270 \pm 0.012$ nm, one can see that this value reduces to 0.228 nm and increases to 0.276 nm at the normal and parallel interface, respectively. This unveils that both in-plane and out-of-plane crystal spacings exhibit distinct responses to the suppressed and promoted Pd diffusion at the interfaces.

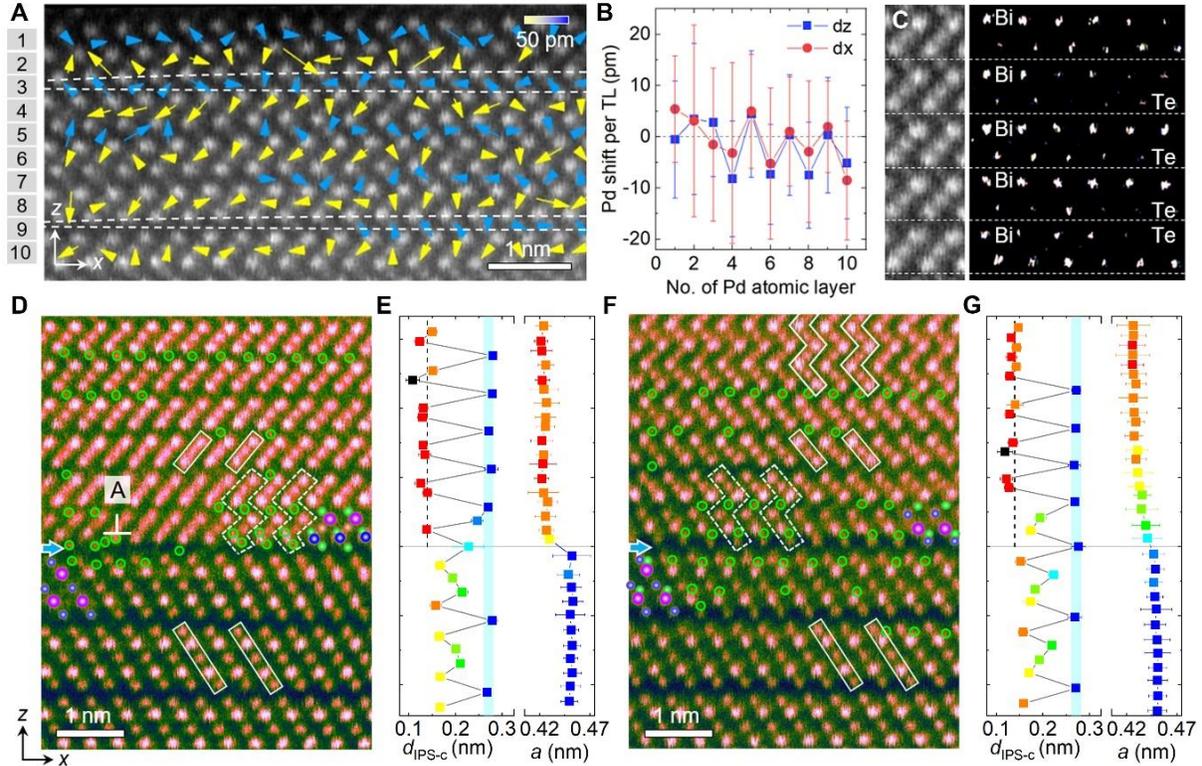

**Figure 2. Pd diffusion induced polarity.** (**A**) HAADF image of xPBT surface with overlapping of Pd displacements. (**B**) Averaged Pd displacements as a function of distance from the surface in the xPBT phase. (**C**) Intensity filtered HAADF image to illustrate the nanoscale ordering of Bi and Te arrangement. (**D,F**) Square-rooted atomic-resolution HAADF images of [100]-oriented xPBT/$Bi_2Te_3$ interfaces with normal and parallel contact, respectively. The discernable Pd columns are annotated by green circles. (**E,G**) Interplanar spacing ($d_{IPS}$) along z and a axis measured across the normal and parallel interfaces, respectively. The extreme values of the vdW spacing is determined by experimental (~0.261 nm) [43] and theoretical (~ 0.281 nm) data reported elsewhere [44,45].

In combination with image simulations with consideration of different Pd occupancy at the (0.0, 0.0, 0.5) site of the $Pd_x(Bi_{0.4}Te_{0.6})_2$ phase, our study reveals that the experimental specimen thickness is around 43.2 nm and the resolvable Pd concentration is ~0.35 (Supplementary Fig. S5). This implies that below this critical concentration, the intercalated Pd atoms within the vdW gaps and in the QLs of the $Bi_2Te_3$ cannot be directly identified. Associated with structural relaxation, our first-principles calculations further verify that the Pd diffusion at the interstitial positions of the $Bi_2Te_3$ is boosted by high-concentration intercalation of Pd atoms at the vdW gaps (Supplementary Figs. S6 and S7). Specifically, instead of the well-defined zigzag structures, the parallel TLs linked by vdW gaps near the xPBT/Bi$_2$Te$_3$ interface play a crucial role in mediating the Pd diffusion across the interfaces, given that the vdW gaps offer large enough space for the dynamic migration of Pd atoms within the lattice matrix (Fig. 2D,F).

To deeply understand the effect of Pd diffusion, relative atomic displacements are mapped near the two xPBT/Bi$_2$Te$_3$ interfaces (Fig. 3A,D). On the xPBT side, by averaging the displacement values of $\delta_{\text{Pd-Bi/Te}}$ along $x$ direction, one can see that the polarity frequently reverses its direction along $z$ direction near the normal interface (Fig. 3B). This gives rise to positively charged head-to-head and negatively charged tail-to-tail vdW interfaces, as schematically illustrated in Fig. 3C. While near the parallel interface, the polarization gradient and head-to-head configuration result in positively charged wall interfaces (Fig. 3E,F). Since the Pd is more electronegative and thus more negatively charged than Bi and Te, this indicates that the parallel interface provides a more favorable condition for the diffusion of Pd into the Bi$_2$Te$_3$ than that at the normal interface. Correspondingly, we see that the in-plane polarity near the normal interface is larger than that near the parallel one, which should be a consequence of local lattice distortion caused by local Pd concentration (Fig. 1I-K).

On the Bi$_2$Te$_3$ side, we find that the Pd self-diffusion leads to development of net out-of-plane polarity in the first few QLs near the interface as well (see empty squares in Fig. 3B,D). This is referred to the primitive QLs, which are composed of octahedral BiTe$_6$ with antiparallel polarity and the total is null in polarity. Near the normal interface, the overall QL polarity points to the -$z$ direction. While near the parallel one, associated with a steady increase of the QL polarity as approaching the interface, the overall QL polarity points to the +$z$ direction. This reveals an interface-dependent switching behavior of polarization, which relates to different Pd diffusion pathways. By correlating with the observed Pd distribution, one can see that the +$z$-oriented QL polarity and head-to-head wall interfaces provide an attractive force for the diffusion of Pd atoms into the TI. Given the demand of polarization screening,[41,46] these results indicate that instead of electronic screening from the metallic xPBT, ionic migration via the Pd diffusion plays a major role in screening the QL polarity in the 2D layered TI. This interprets the suppression and promotion of Pd diffusion at the normal and parallel interface, respectively.

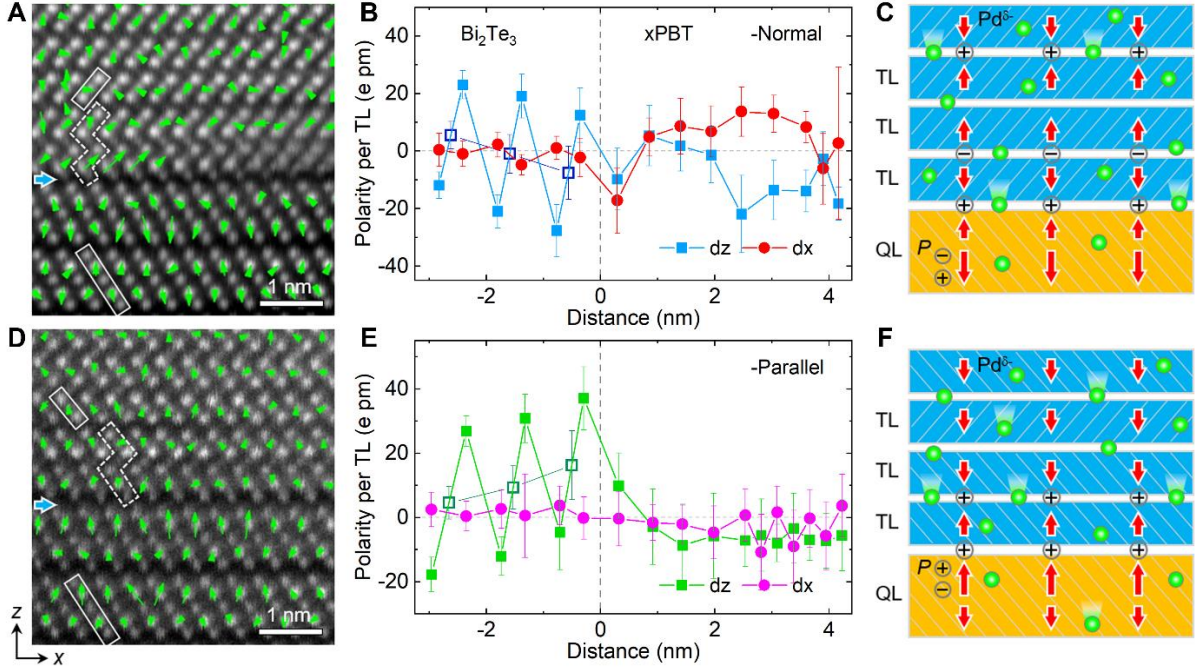

**Figure 3. Polarity-mediated Pd diffusion near the interfaces.** (**A,D**) Mapping of polar displacements of Pd and Bi against centers of their nearest-neighboring Te/Bi ($\delta_{Pd-Bi/Te}$) and Te columns ($\delta_{Bi-Te}$) near the normal and parallel xPBT/Bi$_2$Te$_3$ interface, respectively. (**B,E**) The polarity profiles (by taking the electron charge equals to one) per TL near the normal and the parallel interfaces, respectively. The empty squares denote the polarity per QL in the Bi$_2$Te$_3$. (**C,F**) Illustration of z-axis polarity mediated Pd diffusion near the head-to-head and tail-to-tail dipole interface, respectively. One should note that the electric dipole (pointing from negative to positive charges) direction in the xPBT phase (**A**) and (**D**) is reversed with respect to the polar displacement shown in (**B**) and (**E**), given the larger electronegativity of Pd (denoted by green circles, $\chi_{Pd}$ = 2.20) than that of Te ($\chi_{Te}$ = 2.10) and Bi ($\chi_{Bi}$ = 2.02).

To establish a detailed structure-superconductivity relationship, we perform first-principles calculations using the KKR method[47] on four structural models of the xPBT phases, i.e., pure PdTe and PdTe$_2$ phases, a disordered alloy phase of Pd(Bi$_{0.4}$Te$_{0.6}$)$_x$ and an ordered alloy phase of Pd(BiTe)$_x$ with x = 1 or 2 (Fig. 4A-D). According to the BCS theory, the density of state around the Fermi energy, DOS($E_F$), exponentially influences the superconducting gap and transition temperature. Compared with the normal-state electronic structures, which exhibit large changes between the PdTe and PdTe$_2$ phases, we find that the DOS($E_F$) is reduced by about 25-30% as the PdTe and PdTe$_2$ phases are disordered by random Bi substitution at their Te sites (Fig. 4E,F). Corresponding to an upward shift of the DOS curve, the overall downward shift of $E_F$ thus indicates that the $T_C$ of the xPBT phase is lowered with respect to the PdTe phase.

By taking the similar electron-phonon coupling coefficient ($\lambda$) in PdTe and PdTe$_2$, with $\lambda$ = 0.58 and 0.65,[29,48] our modelling on the intrinsic *s*-wave pairing shows that the superconductivity of the xPBT phase is dominated by electronic degree of freedom and the change in metallicity (Fig. 4G and Supplementary Fig. S8). A detailed analysis on atom- and orbital-resolved contributions indicates that the Pd *d*-electrons are vital to stabilize the superconductivity (Supplementary Fig. S9). On the one hand, with respect to the PdTe$_x$ (x = 1 or 2) phases (see Table S1),[49-52] with a ratio of $T_C$[PdTe$_2$]/$T_C$[PdTe] = 1.7/4.5 = 0.38, its excellent agreement with our calculated ratio about superconducting order parameter $\chi$, $\chi$[PdTe$_2$]/$\chi$[PdTe] = 0.37, unveils that an increasing Bi content tends to reduce the magnitude of $\chi$ and thus the $T_C$ (Fig. 4H and Supplementary Fig. S10). One point worth noting is that an ordered Bi-Te arrangement in Pd$_2$BiTe ($\chi$ = 0.595), introducing an intrinsic polar order in the structure, increases the superconductive $T_C$ with respect to the disordered Pd(Bi$_{0.4}$Te$_{0.6}$) phase ($\chi$ = 0.575), which has less Bi content compared with the Pd(Bi$_{0.5}$Te$_{0.5}$). On the other hand, analogous to the $T_C$ difference between PdTe and PdTe$_2$, a decreasing Pd content also reduces the superconductive $T_C$ as evidenced in the Pd$_2$BiTe and PdBiTe phases. Given that the short-range polar order can be averaged out on the length of tens of nanometers to several micrometers of the superconductive coherence length, we thus speculate that the larger experimental $T_C$ (compared to pure PdTe$_2$) is attributed to mixing of the PdTe- and PdTe$_2$-like phases in xPBT, which results in an effectively larger average superconducting gap.

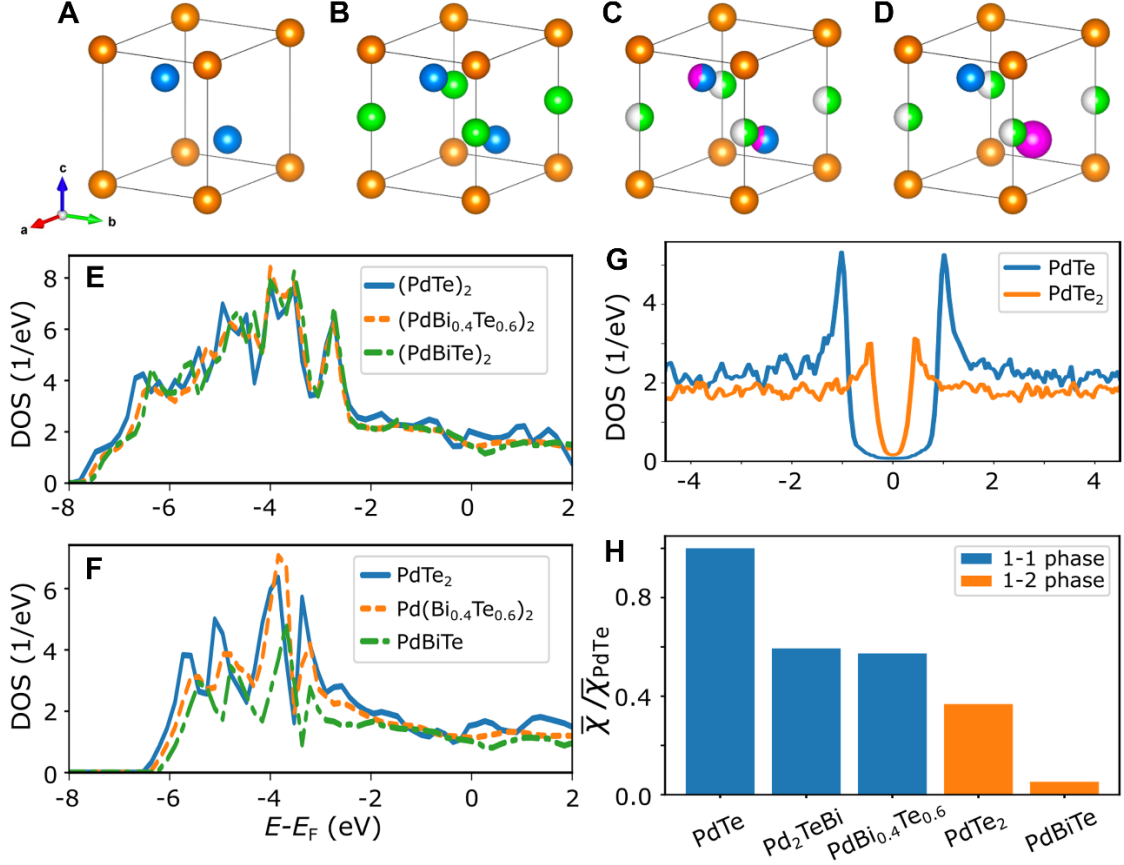

**Figure 4. Electronic structure and superconductivity of the xPBT phase.** (**A-D**) Crystal structure motifs of (**A**) $PdTe_2$, (**B**) PdTe, (**C**) random alloy $Pd_x(Bi_{0.4}Te_{0.6})_2$, (**D**) ordered alloy $Pd_{1+x}(BiTe)_2$ chosen for the DFT calculations. The spheres are color-coded as follows: Pd-orange, Te-blue, Bi-pink, intercalated Pd-green, where half-filled green spheres refer to either presence or absence of Pd in the $PdTe_x$ motif. (**E,F**) Normal state DOS of $(PdBi_{0.4}Te_{0.6})_2$ and $Pd(Bi_{0.4}Te_{0.6})_2$, respectively. (**G**) Superconducting gap in the DOS of PdTe and $PdTe_2$. (**H**) Magnitude of the average superconducting order parameter of different xPBT phases.

On this basis, we further calculate the band structures of relaxed $3 \times 3 \times 1$ $Bi_2Te_3$ supercells with different concentrations of Pd intercalated into the vdW gaps of the TI (Fig. 5A,D and Supplementary Fig. S6). Being consistent with our experimental observation, we find that the high Pd content intercalation mainly leads to structural relaxation along the *z* direction, where the underlying Bi atoms are pushed away from their high symmetry positions and unequal electric dipoles may form within the QL (Supplementary Fig. S7). Further, we investigate the robustness of the topological band inversion of the TI upon increasing Pd concentration. It is found that the Pd *d*-bands form in the bulk bandgap region of $Bi_2Te_3$ (Fig. 5B,C and Supplementary Fig. S11). When Pd atoms diffuse into the QL structure at larger Pd concentration, the Pd *d*-bands move closer to the top of the valence band and a clear bandgap survives. This reveals that the intercalated Pd atoms and the induced polarity may largely modify the band structure of the $Bi_2Te_3$.

As is known, the topological phase transition in $Bi_2Te_3$ happens when the order of Bi *p*- and Te *p*-states become inverted around the Γ point due to the spin-orbit coupling (SOC).[53] Following the Te *p*-character of the bands around Γ upon activating SOC (Supplementary Fig. S12), our calculation proves that the topological phase transition stays intact at low Pd concentration. Although more Pd-derived impurity bands appear within the TI's bandgap, the topological band inversion is preserved even at larger concentration of Pd intercalation into the vdW gap of $Bi_2Te_3$ (Fig. 5E,F). One should note that the flat lines around $E_F$ (highlighted by black arrows in Fig. 5) arise from *d* bands of the intercalated Pd atoms. Although the bands tend to hide the TI's bandgap, they do not break its topological nature. This is inferred from the surviving band inversion visible in the Te *p*-character upon including SOC (black ellipses in Fig. 5B,E). These results of the bulk electronic structure of Pd-intercalated $Bi_2Te_3$ suggest the robustness of topological surface states at the xPBT/$Bi_2Te_3$ interface. Therefore, a topological superconductor can be expected at such a S/TI heterostructure due to the good proximity effect.[4]

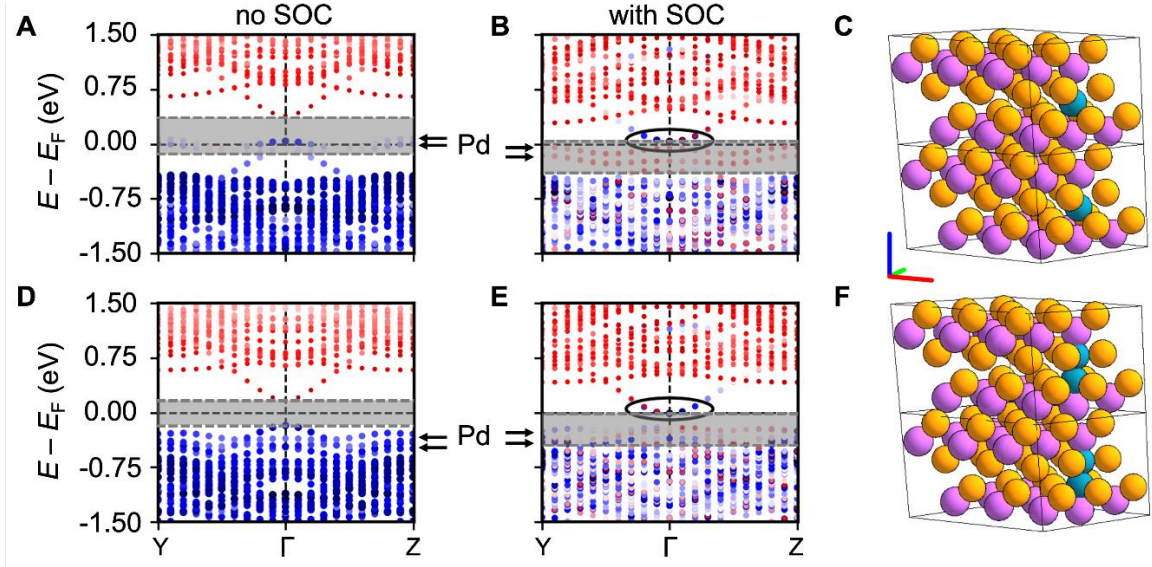

**Figure 5. Band inversion in $Bi_2Te_3$ upon Pd diffusion. (A,B)** Band structures calculated without and with consideration of the SOC for a $3 \times 3 \times 1$ supercell of $Bi_2Te_3$ (Bi-purple, Te-orange) with a single Pd atom intercalated into the vdW gap shown in (**C**). (**D,E**) Band structures without and with SOC for 2 Pd atoms intercalated into the vdW gap shown in (**F**). The red-white-blue coloring of the bands reflects that the TI conduction band is mainly composed of Te-$p$ orbital character. The black ellipses highlight the topological band inversion visible in the transferred orbital character from red to blue in the bottom of the conduction band. The grey shaded areas indicate the TI's bandgap and location of the flat Pd impurity bands is indicated by the black arrows.

One question worth noting is that if the Pd diffusion happening only at the xPBT/$Bi_2Te_3$ interface, will the resulting metallic Pd $d$-states in the bandgap region stand in the way of realizing a topological superconductor? As is known, the recipe for engineering a topological superconductor in such an S/TI heterostructure requires: (i) existence of a topological surface state that can be proximitized; (ii) no other states that are not proximitized and thus close the proximity-induced gap in the electronic structure of the TI. Here we argue that both conditions are met at the xPBT/$Bi_2Te_3$ interface. While there are Pd $d$-derived states in the bandgap of the TI upon Pd diffusion, these are metallic states that are expected to hybridize well with the electronic structure of $PdTe_x$ due to their compatible orbital character. Because the Pd $d$-states in the xPBT phase are decisive in achieving robust superconductivity, one can conjecture that the Pd-$d$ impurity states may give rise to a sizable proximity gap and will thus be gapped out. The robustness of the topological phase transition upon Pd diffusion furthermore suggests that the topological surface state will be present at the xPBT/$Bi_2Te_3$ interface and can be proximitized.

The potential overlap and hybridization with Pd-derived states inside the TI's bulk band gap may even be beneficial for the hybridization of the TI's surface state with the superconductor. This may lead to larger proximity gaps in the topological surface state than without intercalated Pd atoms. Since the observed dislocations are accompanied with accumulation of Pd atoms on the xPBT side, our calculation results (Fig. 4H) show that this should rather strengthen the superconductivity of xPBT in these regions. Furthermore, since the coherence length, ranging from tens of nanometers to a few micrometers for PdTe and $PdTe_2$,[30,50] is much longer than the dislocation-related structural and compositional fluctuation (less than ~10 nm), we argue that there will be an averaging out of the superconductive gap on the length scale of the coherence length. Ultimately this leads to a robust superconducting gap in xPBT that can proximitize the TI surface state for generating the topological superconductivity. In addition, no significant charge transfer between the xPBT and TI phases is observed in our calculated band structures (Supplementary Fig. S12). This indicates that detrimental band bending effects unveiled in our earlier work on Nb/$Bi_2Te_3$ interface[54] are absent, which makes this S-TI interface a good candidate to engineer a topological superconductor.

**CONCLUSIONS**

In summary, our atomic-scale electron microscopy study reveals two distinct interfaces, the normal and parallel ones between the xPBT and $Bi_2Te_3$ phases. On the basis of quantitative image analysis, we find that the inhomogeneous Pd diffusion induces polarization in the xPBT phase and the Pd-intercalated $Bi_2Te_3$ phase, respectively. Specifically, it is found that the Pd diffusion is synergistically controlled by interfacial lattice strain and QL polarity, which inherently couple with the diffusion concentration of Pd atoms near the interfaces. Our first-principles calculations point out that the superconductivity of the xPBT phase is robust against the inversion symmetry breaking and chemical disorder. Although the Pd diffusion breaks the structural symmetry of the $Bi_2Te_3$, the metallic Pd-derived states in the bulk bandgap does not destroy the topological band inversion. These findings not only unravel the diffusion pathway of metals into the 2D layered TIs, which may apply to Nb-, Cu- or Sr-doped $Bi_2Se_3$ with nematic superconductivity,[57,58] but also highlight the necessity of exploring the potential role of electric polarization[59] on electron pairing as studying MZMs in such S-TI heterostructures.

**MATERIALS AND METHODS**

**Thin film growth:** The samples were grown as thin films via molecular-beam epitaxy (MBE). First, $10 \times 10$ mm$^2$ Si(111) samples were prepared by a standard set of wafer cleaning steps (RCA-HF) to remove organic contaminations and the native oxide. A consecutive HF dip passivates the Si surfaces with hydrogen for the transfer into the MBE chamber (base pressure $5 \times 10^{-10}$ mbar). To desorb the hydrogen from the surface, the substrates were heated up to 700 °C for 10 min and finally cooled down to 275 °C. The tellurium shutter was opened several seconds in advance to terminate the silicon surface by Te, which saturates the dangling bonds. Following this, standard Bi and Te effusion cells with vacuum being at $2.2 \times 10^{-8}$ mbar and $5.7 \times 10^{-7}$ mbar were heated to $T_{Bi} = 460$ °C and $T_{Te} = 260$ °C, respectively, for growth of the Bi$_2$Te$_3$ films. After this, the sample was cooled down to -20 °C (in vacuum) and the Pd was deposited via the e-beam evaporation on top of the Bi$_2$Te$_3$.

**Scanning transmission electron microscopy experiments:** For TEM observations, cross-sectional lamella specimens with a dimension of around 4 μm × 10 μm were cut along Si [1$\bar{1}$0] direction using focused ion beam (FIB, FEI Helios NanoLab 400S) system and NanoMill (Model 1040) was used to mill down and remove the surface contamination. An FEI Titan 80-200 ChemiSTEM microscope equipped with a HAADF detector and a Super-X energy-dispersive X-ray spectrometer was used to collect the STEM image and EDX results. With a semi-convergent angle at 24.7 mrad, the HAADF images were collected in an angle range of 70 ~ 200 mrad. The Dr. Probe software package was used for image simulation,[60] and CrystalMaker and VESTA software packages were used for drawing the crystal structures. The lattice parameters of the xPBT and Bi$_2$Te$_3$ phases are measured and calibrated by referring to that of the Si substrate. The result shows that the lattice parameters is $a = 0.425 \pm 0.002$ nm and $c = 0.538 \pm 0.010$ nm for the xPBT phase, and $a = 0.451 \pm 0.001$ nm and $c = 3.013 \pm 0.015$ nm for the Bi$_2$Te$_3$ phase, respectively.

**First-principles calculations:** In our all-electron density functional theory (DFT) calculations we use the full-potential relativistic Korringa-Kohn-Rostoker Green function method (KKR)[61] as implemented in the JuKKR code[62] as well as the full-potential linearized augmented planewave (FLAPW) code FLEUR.[47] The FLEUR code is used for structural relaxations and the KKR method allows to describe random chemical disorder efficiently via the coherent potential approximation (CPA). The JuKKR code also comes with an extension to the Kohn-Sham-Bogoliubov-de Gennes method, which allows to calculate superconducting properties.[54,63] The series of DFT calculations in this study are orchestrated with the help of the AiiDA-KKR[64,65] and AiiDA-FLEUR[66,67] plugins to the AiiDA infrastructure.[68] This has the advantage that the full data provenance (including all values of numerical cutoffs and input parameters to the calculation) is automatically stored in compliance to the FAIR principles of open research data.[69] The complete data set of this project is made publicly available in the materials cloud archive.[70,71]

**Data and code availability:** The source codes of the AiiDA-KKR plugin,[65] the AiiDA-FLEUR plugin,[67] the JuKKR code,[62] and the FLEUR code[47] are published as open source software under the MIT license at https://github.com/JuDFTteam/aiida-kkr, https://github.com/JuDFTteam/aiida-fleur, https://iffgit.fz-juelich.de/kkr/jukkr, and https://iffgit.fz-juelich.de/fleur/fleur, respectively. The AiiDA dataset containing the DFT calculations of this work are published in the materials cloud archive [71]. The experimental data is available upon reasonable request.

## ACKNOWLEDGEMENT

We are grateful to Prof. Björn Trauzettel for fruitful discussions. This project was funded by the Deutsche Forschungsgemeinschaft (DFG, German Research Foundation) under Germany's Excellence Strategy - Cluster of Excellence Matter and Light for Quantum Computing (ML4Q) EXC 2004/1 – 390534769. X.-K.W. also thanks financial supports from the National High-Level Youth Talents Program (Grant No. 0040/X2450224). P.R. thanks the Bavarian Ministry of Economic Affairs, Regional Development and Energy for financial support within the High-Tech Agenda Project "Bausteine für das Quantencomputing auf Basis topologischer Materialien mit experimentellen und theoretischen Ansätzen". We are grateful for computing time granted by the

## Author contributions

X.-K.W. conceived the research idea, performed the STEM experiments and analyzed the data with help of J.M. A.R.J. initiated the research and grew the thin films under the supervision of D.G. P.R. performed the DFT simulations and analyzed the results. S.B., Y.A. and J.M. contributed helpful discussion and useful sample information. X.-K.W. and P.R. wrote and revised the manuscript with comments from all co-authors.

## Notes

The authors declare that they have no competing interests.

# SUPPLEMENTARY MATERIAL

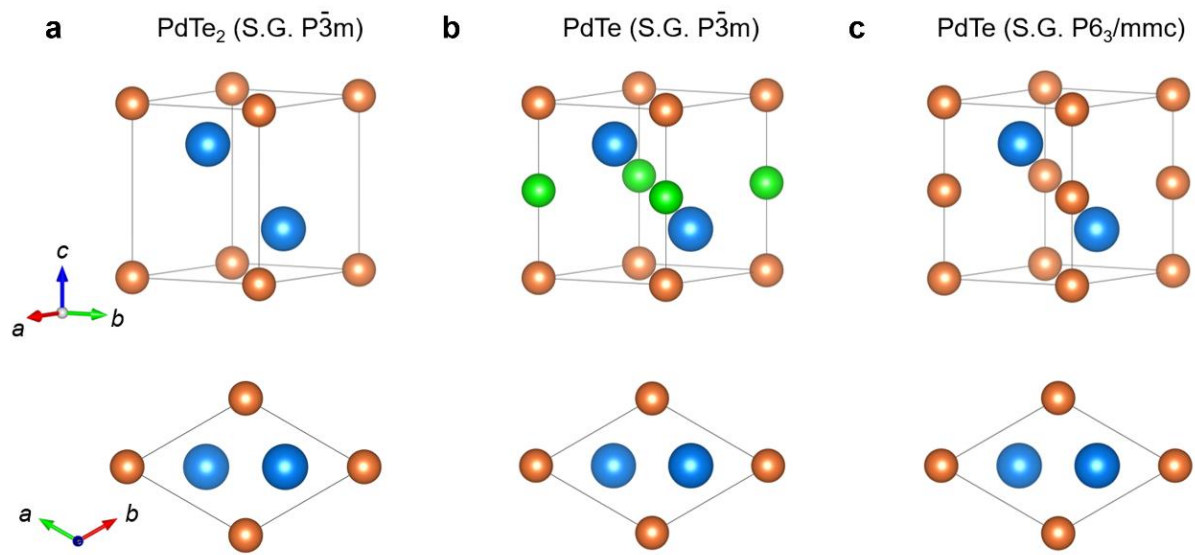

**Figure S1. Comparison between different structural phases. a-c)** Crystal structures of PdTe$_2$ and PdTe with space group (S.G.) P$\bar{3}$m, and PdTe with S.G. P6$_3$/mmc, respectively. The atom types are colored in blue (Te), orange (Pd) and green (intercalated Pd). The top row is the stereo view and the bottom row is the projection view along the *c* axis.

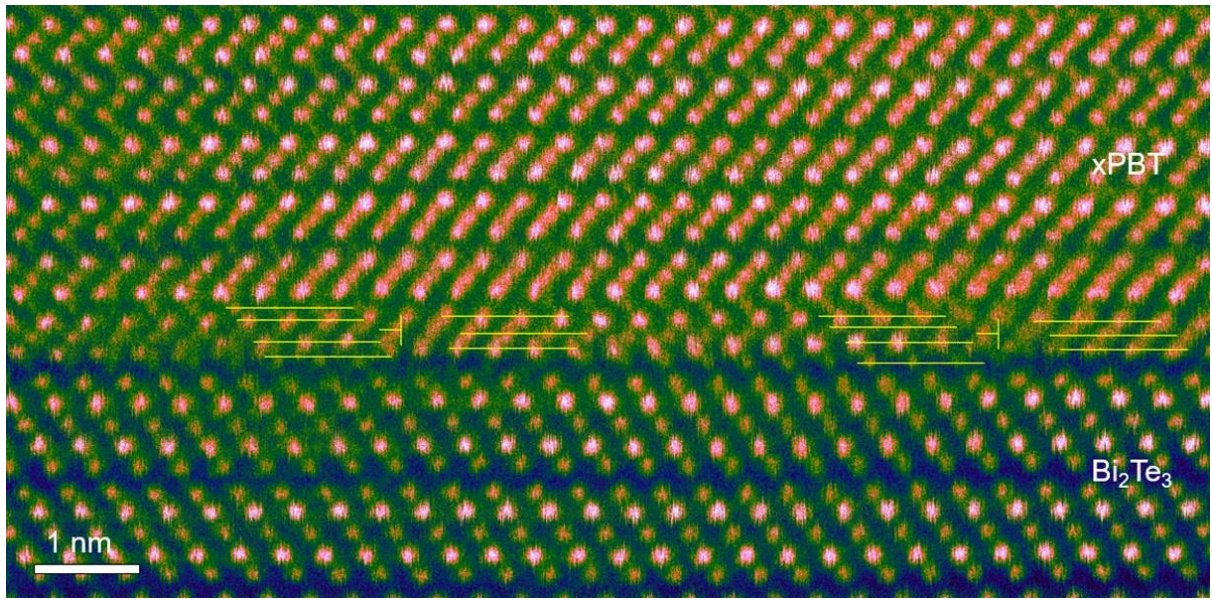

**Figure S2. Atomic-resolution xPBT/Bi₂Te₃ interface.** Yellow solid lines are marked on the HAADF image to show an additional component of the interfacial mismatch dislocation caused by asymmetric Pd accumulation.

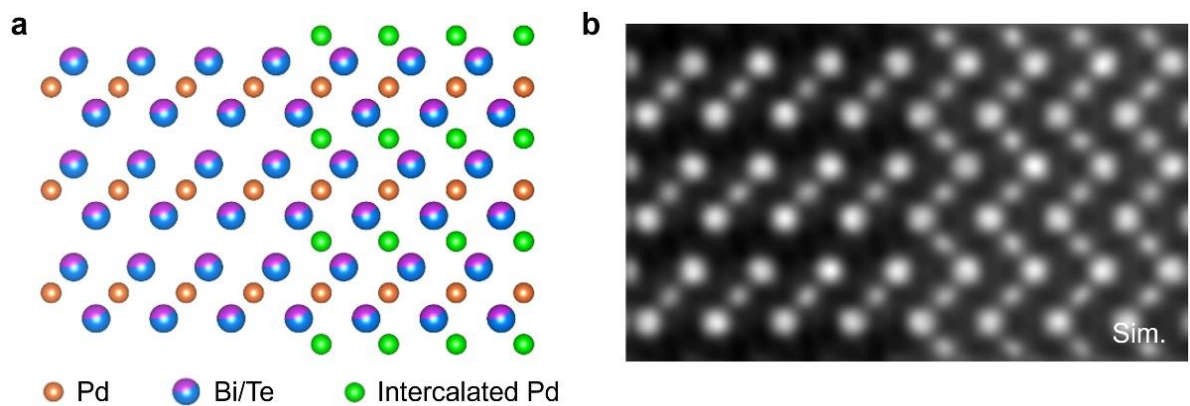

**Figure S3. Image simulation on the xPBT phase. a**) Structural model with and without Pd intercalation. **b**) Simulated HAADF image with a thickness of 43.2 nm.

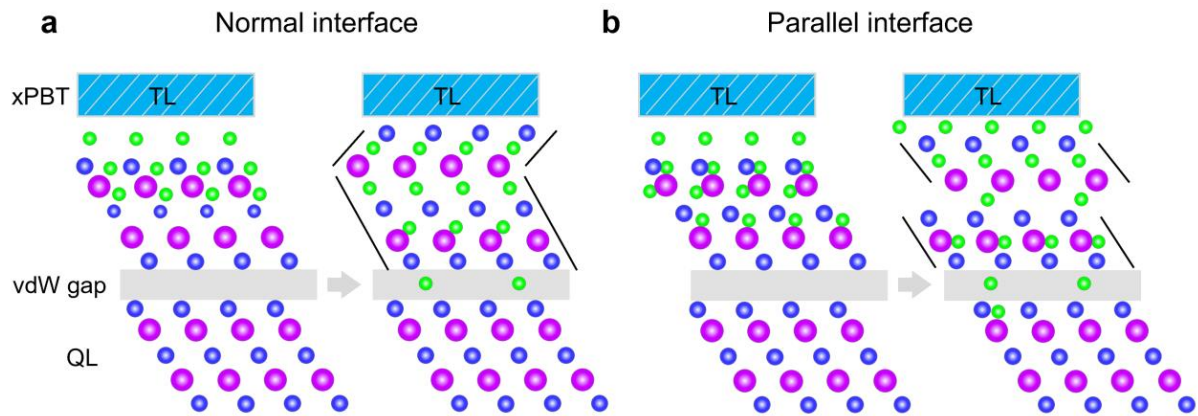

**Figure S4. Schematic Pd diffusion pathways. a,b**) Possible formation of normal-mode and parallel-mode contact interfaces, respectively. Although the breaking of Bi-Te bonds by intercalated Pd atoms dominates in both scenarios, it seems that the intercalation at the vdW gaps may suppress the chemical diffusion, as manifested by the highly condensed $\varepsilon_{yy}$ strain.

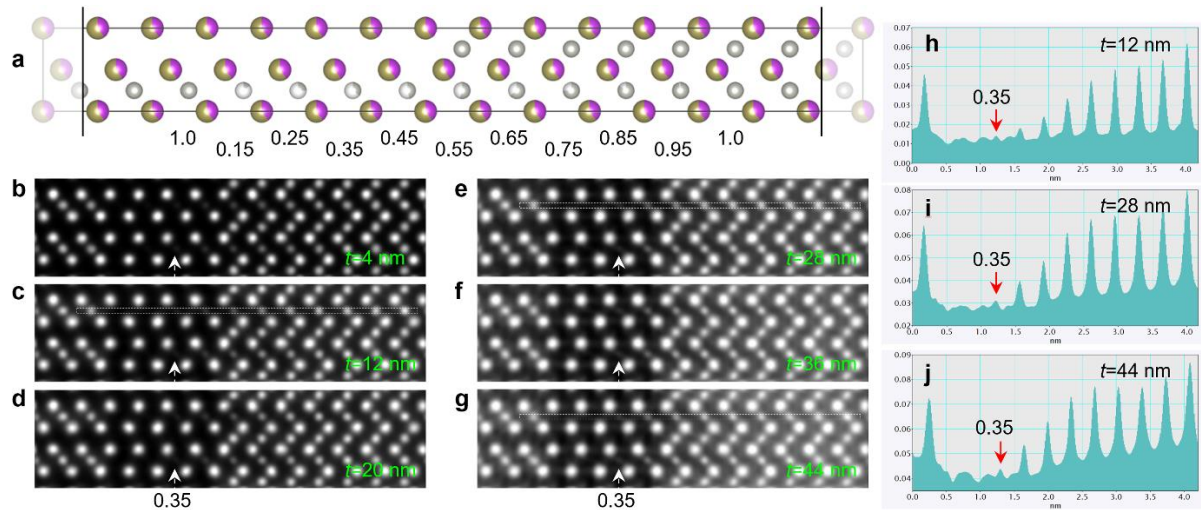

**Figure S5. HAADF image simulation of Pd intercalation with different occupancy at the vdW gaps. a)** Structure model. **b-g)** Simulated HAADF images with thickness ranging from 4 to 44 nm. **h-j)** Intensity profiles extracted from images of c), e) and g), respectively.

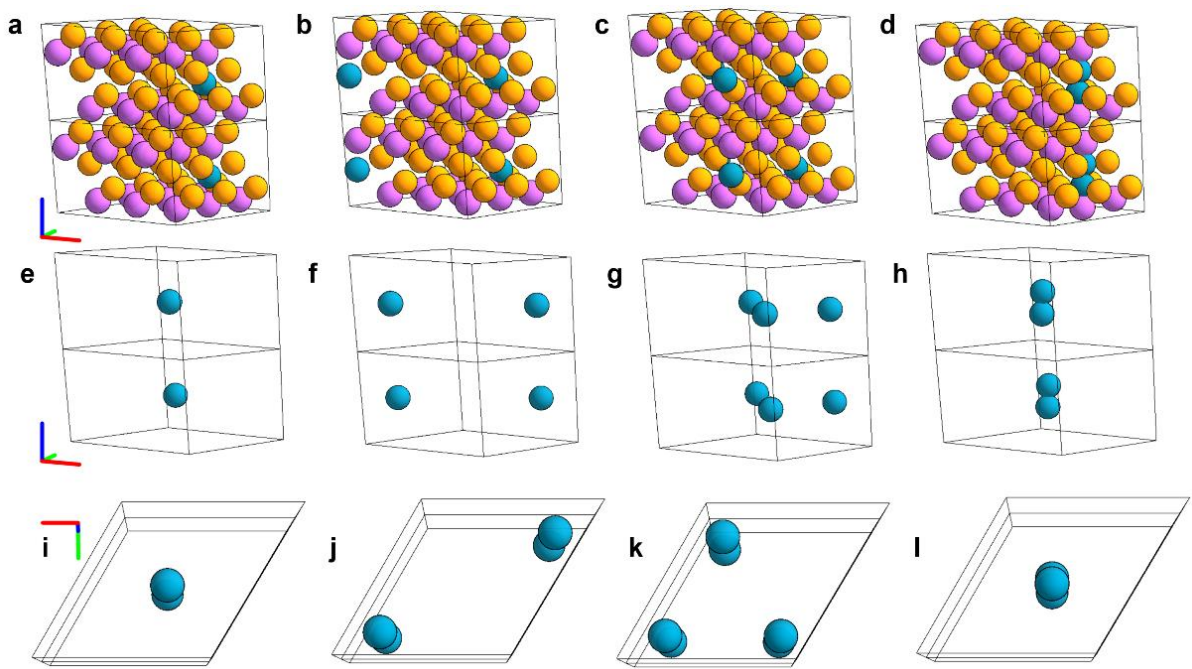

**Figure S6. Relaxed position of Pd atoms intercalated in the vdW gap of Bi$_2$Te$_3$. a-d**) Relaxed positions of one to three Pd atoms in a 3 × 3 × 1 Bi$_2$Te$_3$ supercell, calculated with the FLEUR code. The atom types are Pd (blue), Bi (pink) and Te (orange). **e-h**) Side view of only the Pd atoms, respectively. **i-l**) Top view of the Pd atoms in the unit cells, respectively. When the Pd concentration is low, a-c), the individual Pd atoms remain in the center of the vdW gap. Only when two Pd atoms occupy the same hollow site d), the Pd atoms repel each other and are pushed deeper into the QL, ending up at roughly the same height as the neighboring Te atoms.

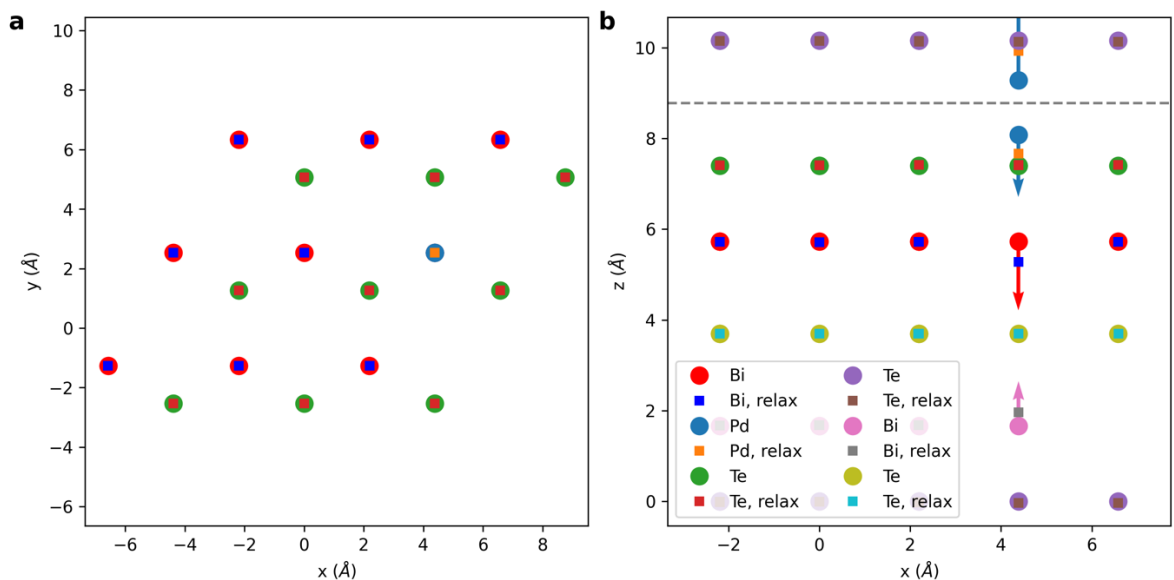

**Figure S7. Broken symmetry upon Pd interaction in Bi$_2$Te$_3$. a)** [001] view of the Pd-intercalated Bi$_2$Te$_3$. **b)** [1$\bar{1}$0] view showing the relaxation of Bi$_2$Te$_3$ with two Pd atoms intercalated at the vdW (grey dashed line) gap as illustrated in Fig. S6d. The original and relaxed positions are indicated by colored spheres and squares, respectively. The direction of the atom relaxation is highlighted by color arrows. One can see that the atoms relax only considerably along *z*-direction of Bi$_2$Te$_3$. With inward movement of Pd atoms into the outermost Te layer of the Bi$_2$Te$_3$ quintuple layer (QL), the underlying Bi atoms (red circle with arrow and relaxed blue square) are attracted towards the center of the QL.

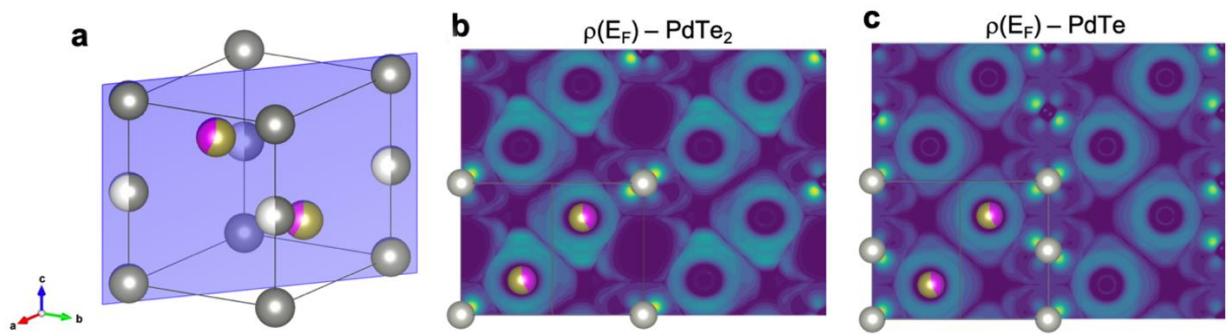

**Figure S8. Electron density distribution in PdTe and PdTe$_2$. a**) Unit cell of PdTe$_x$ where the blue shaded surface highlights the plane in which the charge density at the Fermi level is visualized for **b)** PdTe$_2$ and **c)** PdTe. The additional Pd atoms in c) add additional states (zig-zag pattern in vertical direction along the line formed by Pd atoms) that leads to an increased metallicity and enhanced superconductivity in PdTe compared to PdTe$_2$.

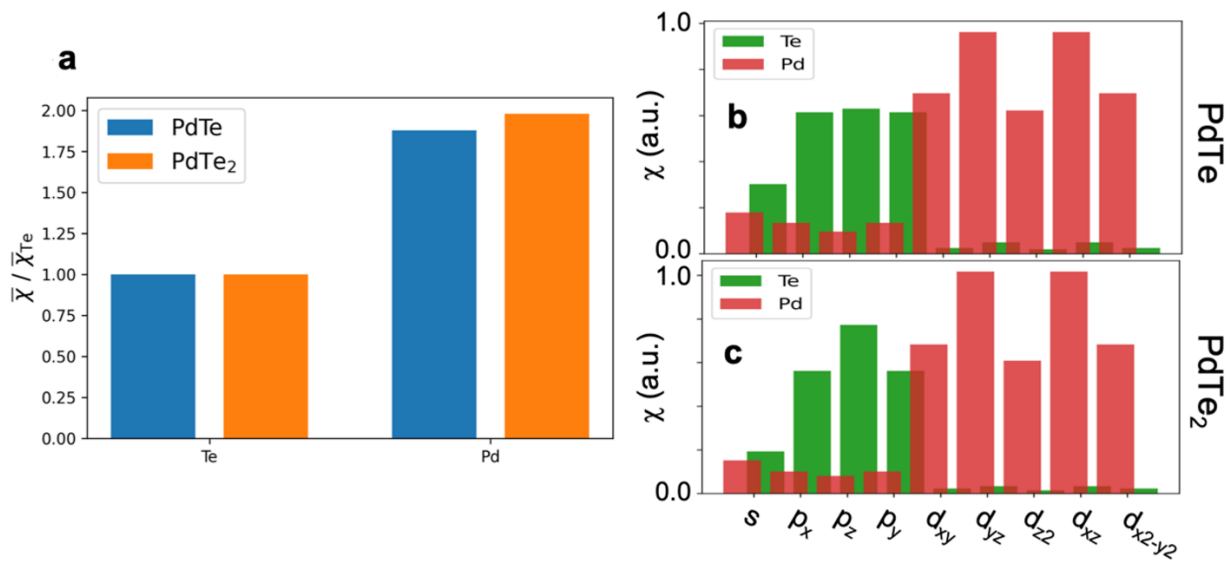

**Figure S9. Atom and orbital resolved anomalous density of PdTe and PdTe$_2$. a)** Atom resolved anomalous density for PdTe (blue) and PdTe$_2$ (orange) relative to the contribution of one Te atom. **b,c)** Atom and orbital resolved anomalous density for PdTe (b) and PdTe2 (c).

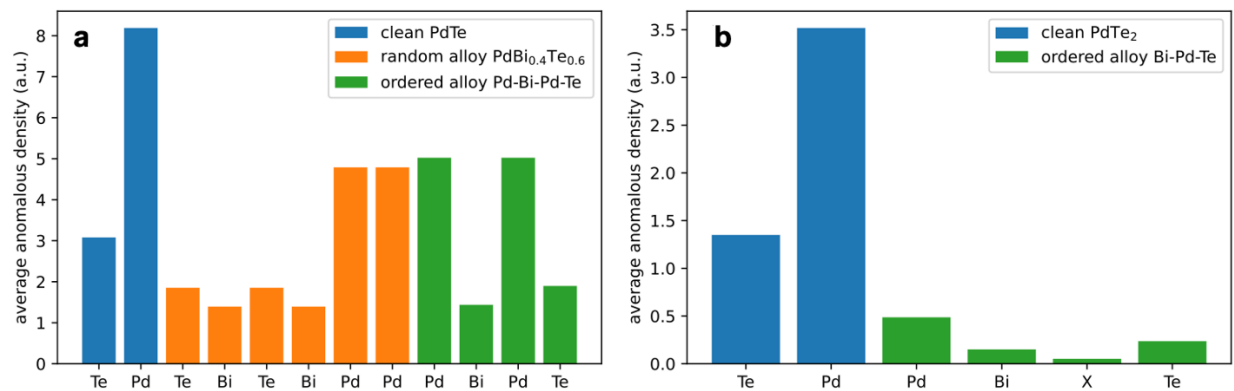

**Figure S10. Atom resolved anomalous density in different xPBT phases. a**) Atom-resolved anomalous density in clean PdTe (blue), the random alloy PdBi$_{0.4}$Te$_{0.6}$ and in the ordered alloy Pd$_2$BiTe. **b)** Atom resolved anomalous density in clean PdTe$_2$ and the ordered alloy phase of PdBiTe.

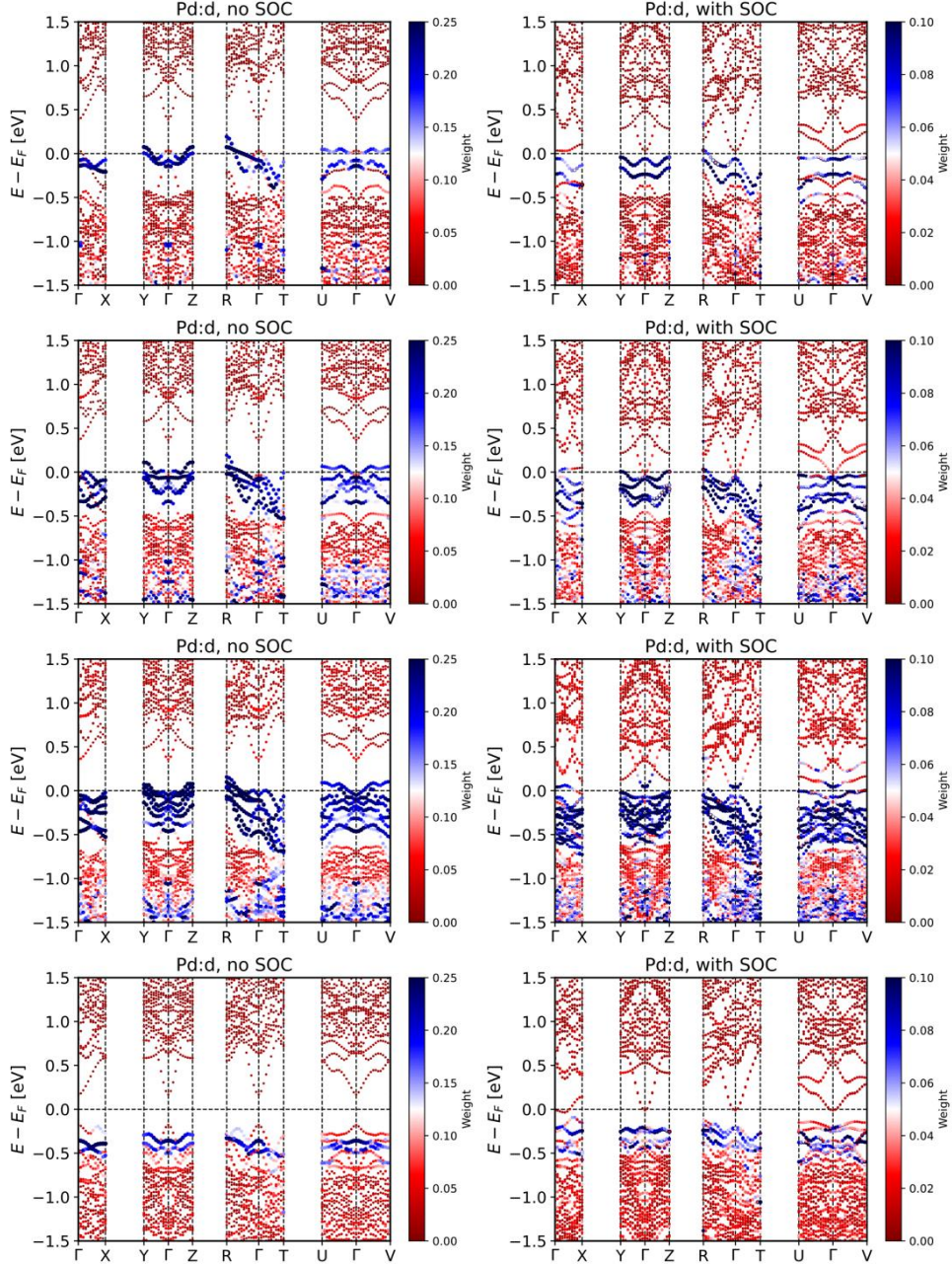

**Figure S11. Contribution of Pd *d* orbitals to the band structure of Pd-intercalated Bi₂Te₃.** The four rows correspond (from top to bottom) to the structural models given in Fig. S6 (a-d) for one to three Pd atoms intercalated into the vdW gap of a 3 × 3 × 1 Bi₂Te₃ supercell. The last row corresponds to the experimentally observed position of Pd atoms at larger concentration. The Pd *d*-states are highlighted in blue. Left column: band structure without SOC, right column: band structure including SOC.

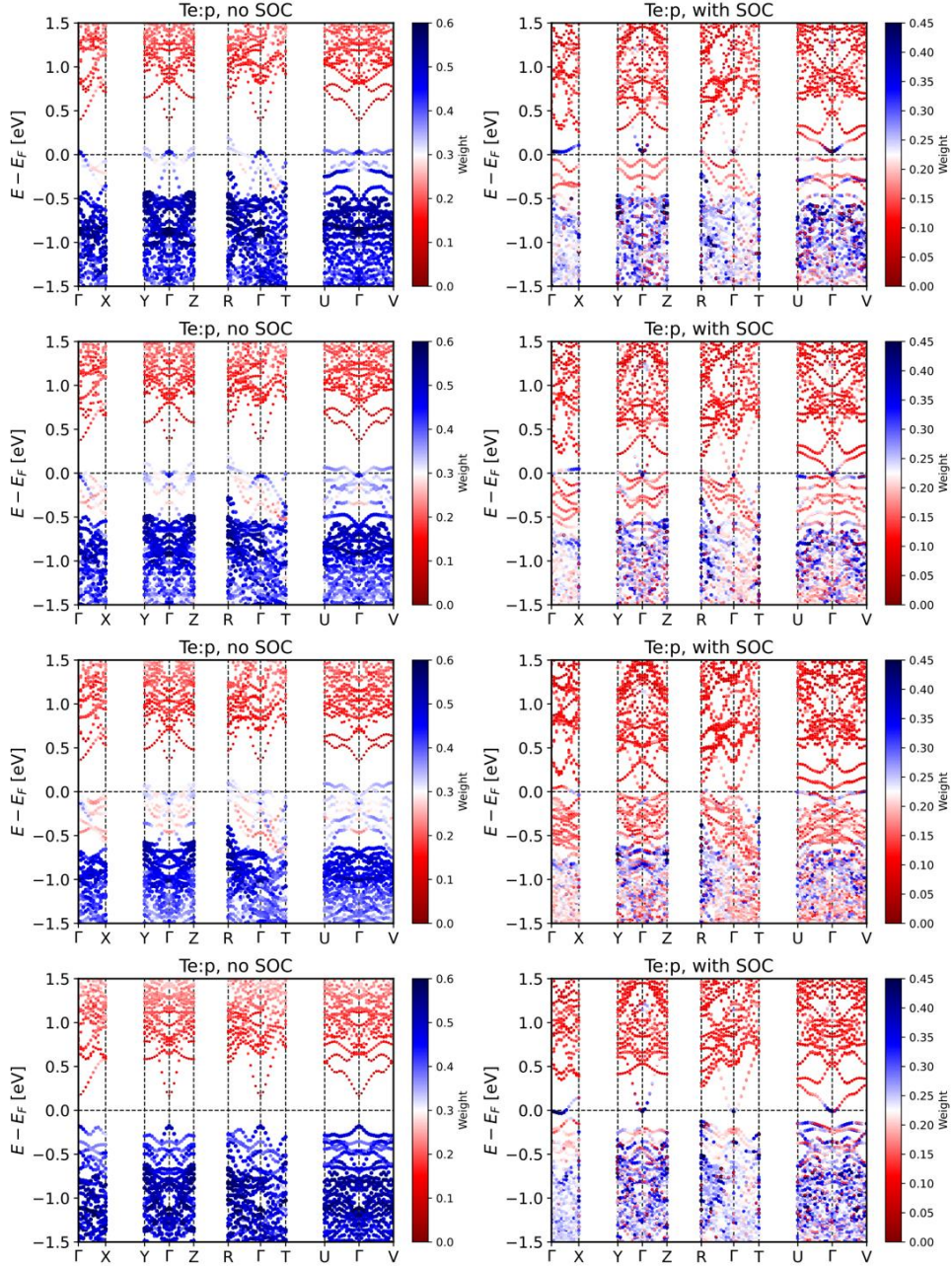

**Figure S12. Contribution of Te *p* orbitals to the band structure of Pd-intercalated Bi$_2$Te$_3$.** Layout of the plots as in Figure S8 but the blue colors highlight the Te *p*-states for the four structural model given in Fig. S6(a-d). Without SOC (left column) the valence band is made up of predominantly Te *p*-states whereas the conduction band (CB) is of predominantly Bi *p* character. With SOC (right column) the topological band inversion is visible around Γ as the blue color at the bottom of the CB. This is only weakened in the third row where many Pd *d*-states are present inside the Bi$_2$Te$_3$ in the band gap (cf. Fig. S10).

**Table S1.** Experimental superconducting transition temperatures $T_C$ of different PdTe$_x$ compounds.

| Material | $T_C$ [K] | References |
|---|---|---|
| PdTe | 4.5 | Refs. 30, 31 |
| PdTe$_2$ | ~1.7 | Refs. 28, 29, 50, 51 |
| Pd(Te,Se)$_2$ | 2.74 | Ref. 52 |
| Cu-intercalated PdTe$_2$ | 2.4 | Ref. 49 |
| Pd-Bi$_2$Te$_3$ | 0.67-1.22 | Ref. 33 |